\documentclass[aps,pre,graphicx,groupedaddress,reprint,showpacs]{revtex4-1}
\usepackage{amsmath}
\usepackage{mathtools}
\usepackage{amssymb}
\usepackage{graphicx}
\usepackage{hyperref}
\usepackage{tensor}
\usepackage{txfontsb} 
\usepackage{multirow}
\DeclareMathAlphabet{\mathcal}{OMS}{cmsy}{m}{n} 
\hypersetup{
    colorlinks=true,
    linkcolor=blue,
    citecolor=blue,
    filecolor=blue,
    urlcolor=blue,
}
\usepackage{listings}

\DeclareSymbolFont{largesymbolsCM}{OMX}{cmex}{m}{n}

\let\sum\relax
\DeclareMathSymbol{\sum}{\mathop}{largesymbolsCM}{"50}
\DeclareSymbolFont{largesymbolsCM}{OMX}{cmex}{m}{n}

\let\prod\relax
\DeclareMathSymbol{\prod}{\mathop}{largesymbolsCM}{"51}
\DeclareSymbolFont{largesymbolsCM}{OMX}{cmex}{m}{n}

\let\intop\relax
\DeclareMathSymbol{\intop}{\mathop}{largesymbolsCM}{"52}

\let\int\relax
\def\int{\intop\nolimits}

\begin{document}

\title{ \ \\ Minimax rational approximation of the Fermi-Dirac distribution}

\author{Jonathan E. Moussa}
\email{godotalgorithm@gmail.com}
\affiliation{Center for Computing Research, Sandia National Laboratories, Albuquerque, New Mexico 87185, USA}

\begin{abstract} 
\bigskip
 \centerline{\begin{minipage}{0.79\textwidth}
\ \ \ \ Accurate rational approximations of the Fermi-Dirac distribution are a useful component in many numerical algorithms for electronic structure calculations.
The best known approximations use $O( \log (\beta \Delta) \log (\epsilon^{-1}))$ poles to achieve an error tolerance $\epsilon$ at temperature $\beta^{-1}$ over an energy interval $\Delta$.
We apply minimax approximation to reduce the number of poles by a factor of four and replace $\Delta$ with $\Delta_{\mathrm{occ}}$, the occupied energy interval.
This is particularly beneficial when $\Delta \gg \Delta_{\mathrm{occ}}$, such as in electronic structure calculations that use a large basis set.
\bigskip
\end{minipage}}
\end{abstract}

\maketitle

Evaluation of the Fermi-Dirac distribution of a Hermitian matrix is a common component of many electronic structure calculations.
A direct evaluation method is to decompose the matrix into its eigenvectors and real eigenvalues and evaluate the distribution for each eigenvalue.
However, it can be more efficient to approximate the distribution with a polynomial or rational function and then work with matrix polynomials or matrix inverses.
This is the foundation for methods that avoid the cubic-scaling bottleneck of eigenvalue decomposition
 and achieve a quadratic \cite{quadratic_scaling} or linear scaling \cite{linear_scaling} of computational cost with system size.
Since the Fermi-Dirac distribution has poles near the real line, a rational approximation is capable of higher accuracy
 than a polynomial approximation when both have the same number of free parameters.
Another advantage of rational approximations is their decomposition into matrix inverses that can be reused
 in many-body calculations based on single-electron Green's functions \cite{GW_method,RPA_method,RPA_method2}.

Rational approximation of the Fermi-Dirac distribution was first developed as numerical quadrature for contour integrals \cite{Goedecker_contour}.
Recent work has reduced quadrature errors and identified error bounds \cite{Lin_contour}.
A natural terminus of this research problem is an optimal rational approximation with respect to a suitable error metric.
Since the eigenvalue spectrum of the underlying Hermitian matrix is unknown and discrete,
 minimax error is the metric most directly applicable to an error analysis.
Even a modest improvement may have significant future impact if these approximations become widely adopted in some future generation of computational materials science software.

In this paper, we develop a procedure for minimax rational approximation of the Fermi-Dirac distribution.
Optimization of these approximations is both nonlinear and ill-conditioned, especially in the high-accuracy limit.
We mitigate the effects of these numerical problems
 with a continuous evolution of solutions from a limit with an analytic solution.
We provide a tabulation of optimized rational approximations that covers a wide parameter regime
 and the source code to reproduce and extend this tabulation \cite{arXiv_supplemental_info}.
We characterize all approximation errors and relate them to known error bounds.
This provides users with practical error estimates that are tighter than known error bounds but with a similar parameter dependence.

\begin{figure}[b]
\includegraphics{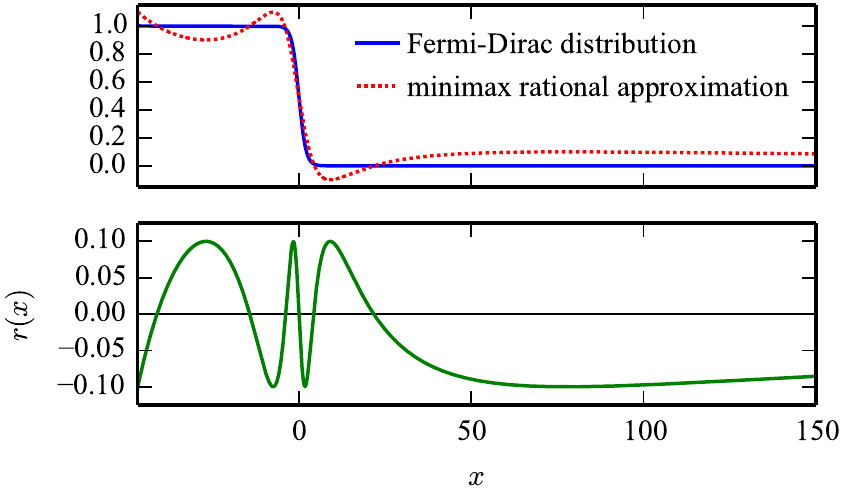}
\caption{\label{fig_example} Simple example of minimax rational approximation of the Fermi-Dirac distribution for $n=3$, $y \approx 46.8$, and $\epsilon_n(y) = 0.1$ (top).
The residual function $r(x)$ equioscillates with 7 extrema inside of the domain of approximation, $[-y,\infty)$ (bottom).
}
\end{figure}

The Fermi-Dirac distribution is defined as
\begin{equation}
 \frac{1}{1 + \exp(\beta (E - E_F))}
\end{equation}
 for temperature $\beta^{-1}$, orbital energy $E$, and Fermi energy $E_F$, all in units of energy.
We seek to approximate the distribution over an interval $E \in [E_{\mathrm{min}},\infty)$
 using a rational function with $n$ poles.
The dimensionless form of the minimax problem is
\begin{equation} \label{minimax}
 \epsilon_n(y) = \min_{w_i, z_i} \max_{x \in [-y,\infty)} \left| \frac{1}{1 + \exp(x)} - \sum_{i=1}^n \frac{w_i}{x - z_i} \right|
\end{equation}
 for $y = \beta (E_F - E_{\mathrm{min}})$.
Here $w_i$ and $z_i$ are the residues and poles of the rational approximation.
For even $n$, the optimal $(w_i, z_i)$ group into conjugate pairs.
For odd $n$, there is one $(w_i, z_i)$ that is real and unpaired with $z_i < -y$.
The residual function,
\begin{equation}
 r(x) = \frac{1}{1 + \exp(x)} - \sum_{i=1}^n \frac{w_i}{x - z_i},
\end{equation}
 for a minimax rational approximation equioscillates between $\epsilon_n(y)$ and $-\epsilon_n(y)$ at $2n+1$ points, as shown in Fig. \ref{fig_example}.
Rational approximations are asymptotically exact in the $x \rightarrow \infty$ limit.

The conventional Remez algorithm \cite{Remez} cannot solve Eq. (\ref{minimax}) for large $n$ or small $\epsilon_n(y)$.
It represents rational functions as a ratio of polynomials, which is a poor numerical representation of highly clustered roots and poles.
Any small perturbation in polynomial coefficients caused by finite-precision arithmetic can cause large perturbations to roots and poles.
The rational functions that approximate the Fermi-Dirac distribution have poles clustered near the origin, which amplify this instability.
We instead develop a method based on a residue-pole form for rational functions that is more numerically stable.

There is a minimax rational approximation problem with an analytic solution \cite{Zolotarev} that is similar to Eq. (\ref{minimax}),
\begin{equation} \label{sgn_fit}
 \tilde{\epsilon}_n(k) = \min_{W_i, Z_i} \max_{X \in [ -1 , -k] \cup [ k, 1]} \left| \mathrm{sgn}(X) - \sum_{i=1}^{n} \frac{W_i}{X - Z_i} \right|
\end{equation}
 for $\mathrm{sgn}(x) = x/|x|$.
We use this result to construct approximate solutions for Eq. (\ref{minimax}).
The similarity between these problems is more apparent after applying a M\"{o}bius transformation,
\begin{equation} \label{mobius}
 x  = - \delta \frac{1 + X d}{X + d}, \ \ \mathrm{where} \ \ \delta = y \frac{ k + d }{ 1 + k d }
\end{equation}
 and $d$ is the zero of the residual of Eq. (\ref{sgn_fit}) that is closest to $k$.
This transformation maps $k \rightarrow -y$, $\pm 1 \rightarrow \mp \delta$, and $-d \rightarrow \infty$.
The transformed and rescaled version of Eq. (\ref{sgn_fit}) is
\begin{equation} \label{step_fit}
  \tfrac{1}{2} \tilde{\epsilon}_n(k) = \min_{w_i, z_i} \max_{x \in [ -y , -\delta] \cup [ \delta, \infty)} \left| \theta(x) - \sum_{i=1}^{n} \frac{w_i}{x - z_i} \right|
\end{equation}
 for the step function $\theta(x) = \tfrac{1}{2} - \tfrac{1}{2} \mathrm{sgn}(x)$.
 
Because of the similarity between target functions,
\begin{equation}
 \max_{x \in [ -y , -\delta] \cup [ \delta, \infty)} \left| (\exp(x) + 1)^{-1} - \theta(x) \right| \le \exp(-\delta) ,
\end{equation}
 optimal solutions of Eq. (\ref{step_fit}) approximate solutions of Eq. (\ref{minimax}).
We choose $y$ to set $\delta = - \ln( \tfrac{1}{4}\tilde{\epsilon}_n(k))$ in Eq. (\ref{mobius}),
 which is large enough to preserve the oscillations of $r(x)$.
It has $2n-2$ local extrema that are close to $\tfrac{1}{2}\tilde{\epsilon}_n(k)$ in magnitude
 and two local extrema in $(-\delta,\delta)$ that can be much larger.
We order the $2n$ local extrema $x_i$ as $x_i < x_{i+1}$.
The minimax solution balances the magnitudes of the local extrema to an error $\epsilon = \tfrac{1}{2} \tilde{\epsilon}_n(k)$,
\begin{equation} \label{equioscillation}
 r(x_i) = (-1)^{i+n} \epsilon \ \ \mathrm{and} \ \ r'(x_i) = 0.
\end{equation}
If the first condition is not satisfied, then we iteratively refine the solution by solving for its linear response,
\begin{equation} \label{linear_response}
 r(x_i) - \sum_{j=1}^n \frac{\delta w_j}{x_i - z_j} - \sum_{j=1}^n \frac{w_j \delta z_j}{(x_i - z_j)^2} = (-1)^{i+n} \epsilon ,
\end{equation}
 and calculating the linear response of the local extrema,
\begin{equation} \label{extrema_response}
 r''(x_i) \delta x_i = - \sum_{j=1}^n \frac{\delta w_j}{(x_i - z_j)^2} - 2 \sum_{j=1}^n \frac{w_j \delta z_j}{(x_i - z_j)^3} .
\end{equation}
The exact local extrema $x_i$ are recomputed after any update of $r(x)$ to maintain the simple form of the linear response.

The analytic solution \cite{Zolotarev} to Eq. (\ref{sgn_fit}) contains Jacobi elliptic functions, $\mathrm{sn}(u,k)$, $\mathrm{cn}(u,k)$, and $\mathrm{dn}(u,k)$,
 and their associated quarter periods, $K(k)$ and $i K' (k)$.
Using convenient variables,
\begin{align}
 k' &= \sqrt{1 - k^2} , & \theta &= \frac{K'(k)}{n} , \notag \\
 \kappa_m &= \frac{k}{\mathrm{dn}(m \theta,k')} , & \lambda_m &= k \frac{\mathrm{sn}(m \theta,k')}{\mathrm{cn}(m \theta,k')} ,
\end{align}
 and a common rational function primitive,
\begin{equation}
 R_n(x) = \frac{1}{x}  \prod_{m=1}^{\lceil(n - 1)/2\rceil} (x^2 + \lambda_{2m - 1}^2) \prod_{m=1}^{\lfloor(n - 1)/2\rfloor}(x^2 + \lambda_{2m}^2)^{-1} ,
\end{equation}
the minimax rational approximations in root-pole form are
\begin{align}\label{root_pole}
  \frac{2 R_{n}(X)^{-1}}{R_{n}(\kappa_1)^{-1} + R_{n}(k)^{-1}} \ \ & \mathrm{for \ even} \ n \notag \\
\mathrm{and} \ \ \frac{2 R_{n}(X)}{R_{n}(k) + R_{n}(\kappa_1)} \ \ & \mathrm{for \ odd} \ n.
\end{align}
Their maximum pointwise error in approximating $\mathrm{sgn}(X)$ is
\begin{align}
 \epsilon_{n}(k) = \left\{ \begin{array}{ll} \dfrac{R_{n}(\kappa_1)^{-1} - R_{n}(k)^{-1}}{R_{n}(\kappa_1)^{-1} + R_{n}(k)^{-1}} , & n \ \mathrm{even} \\[9pt]
   \dfrac{R_{n}(k) - R_{n}(\kappa_1)}{R_{n}(k) + R_{n}(\kappa_1)} , & n \ \mathrm{odd} \end{array} \right.
\end{align}
 which is attained at $X \in \{ \pm k, \pm \kappa_i, \pm1 \}$.
This is bounded by
\begin{equation} \label{analytic_bound}
 \tilde{\epsilon}_n(k) \le 4 \exp\left(- n \frac{\pi K(k)}{K'(k)} \right) \le  4 \exp\left(- n \frac{\pi^2/2}{\ln(4/k)} \right).
\end{equation}
The first inequality is asymptotically tight for large $n$, and the second inequality is asymptotically tight for small $k$.

We use partial fraction decompositions to convert Eq. (\ref{root_pole}) into residue-pole form as in Eq. (\ref{sgn_fit}).
For even $n$, this is
\begin{align}
 W_{2m-1} = W_{2m} &= \frac{1}{R_n(\kappa_1)^{-1} + R_n(k)^{-1}}
  \prod_{p=1}^{n/2} \frac{\lambda_{2m-1}^2 - \lambda_{2p}^2}{\lambda_{2m-1}^2 - \lambda_{2p-1}^2 \overline{\delta}_{m,p} } , \notag \\
  Z_{2m-1} = -Z_{2m} &= i \lambda_{2m-1} .
\end{align}
For odd $n$, the root-pole form of the minimax solution is
\begin{align}
 W_1 &= \frac{2}{R_n(k) + R_n(\kappa_1)} \prod_{m=1}^{(n-1)/2} \frac{\lambda_{2m-1}^2}{\lambda_{2m}^2} , \notag \\
 W_{2m} = W_{2m+1} &= \frac{1}{R_n(k) + R_n(\kappa_1)} \prod_{p=1}^{(n-1)/2} \frac{\lambda_{2m}^2 - \lambda_{2p-1}^2}{\lambda_{2m}^2 - \lambda_{2p}^2 \overline{\delta}_{m,p} } , \notag \\
 Z_1 &= 0 , \ \ \ Z_{2m} = -Z_{2m+1} = i \lambda_{2m} .
\end{align}
Here $\overline{\delta}_{m,p} = 1 - \delta_{m,p}$ is a complementary Kronecker delta.
The transformed minimax solutions for this problem are
\begin{equation}
 w_i = \frac{W_i \delta}{2} \frac{ 1 - d^2 }{ (Z_i + d)^2 } , \ \ z_i =  - \delta \frac{1 + Z_i d}{Z_i + d} .
\end{equation}
This is our initial approximation for the solution to Eq. (\ref{minimax}).

We analytically solve Eq. (\ref{linear_response}) by exploiting its underlying Cauchy matrix structure.
Eq. (\ref{linear_response}) is the $\delta x \rightarrow 0$ limit of
\begin{equation}
 \frac{1}{2} \sum_{j=1}^n \sum_{\sigma \in \{+1,-1\}} \frac{\delta w_j + \sigma w_j \delta z_j / \delta x}{x_i - z_j - \sigma \delta x} = r(x_i) - (-1)^{i+n} \epsilon .
\end{equation}
We then solve this linear system using Cramer's rule and the analytic formula for Cauchy matrix determinants \cite{Cauchy_determinant},
\begin{align}
 p_i &= \prod_{j=1}^{2n} (x_j - z_i) \prod_{j=1 , \ j \neq i}^n {(z_j - z_i)^{-2}}, \notag \\
 q_i &= p_i \left[ \sum_{j=1 , \ j \neq i}^n \frac{2}{z_j - z_i} - \sum_{j=1}^{2n} \frac{1}{x_j - z_i} \right] , \notag \\
 s_i &= \prod_{j=1}^n (z_j - x_i)^2 \prod_{j=1 , \ j \neq i}^{2n} {(x_j - x_i)^{-1}}, \notag \\
 \delta w_i &= \sum_{j=1}^n \left[ \frac{p_i s_j}{(x_j - z_i)^2} + \frac{q_i s_j}{x_j - z_i} \right] \left[ r(x_j) - (-1)^{j+n} \epsilon \right] , \notag \\
 \delta z_i &= \frac{1}{w_i} \sum_{j=1}^n \frac{p_i s_j}{x_j - z_i} \left[ r(x_j) - (-1)^{j+n} \epsilon \right] .
\end{align}
We update solutions as $w_i \rightarrow w_i + \eta \, \delta w_i$, $z_i \rightarrow z_i + \eta \, \delta z_i$, and $x_i \rightarrow x_i + \eta \, \delta x_i$
 for $\eta \in (0,1]$ that is small enough to preserve oscillations of $r(x_i)$.
To mitigate detrimental nonlinear effects when $\epsilon \ll 1$,
 we continuously evolve solutions at fixed $n$ from large $\epsilon$ values to small $\epsilon$ values.

We tabulate \cite{arXiv_supplemental_info} the minimax rational approximations of the Fermi-Dirac distribution for $n$ poles and accuracy $\epsilon = 10^{-m}$
 with $1 \le n \le 100$ and $2 \le m \le 13$.
This range of accuracy is limited by the use of double-precision floating-point numbers.
More accurate approximations require arithmetic with higher precision.
The tabulation is also restricted to $y \ge 100$ because smaller $y$ values have limited practical value.
For $y \ge 10$, the error in all computed solutions is empirically bounded by
\begin{equation} \label{empirical_error}
 \epsilon_n(y) \le 2 \exp\left( - n \frac{\pi^2/2}{\ln(\pi y)} \right),
\end{equation}
 using a parameterized form similar to Eq. (\ref{analytic_bound}).

The energy scales of interest in typical electronic structure calculations are usually given in units of electron volts (eV).
These include the temperature $\beta^{-1}$, the bandwidth of occupied electrons, $\Delta_{\mathrm{occ}} = E_F - E_{\min}$,
 and the full electronic bandwidth of the simulation, $\Delta = E_{\max} - E_{\min}$.
Typical calculations use near-ambient temperatures, $\beta^{-1} \sim 0.01$ eV,
 and consider only valence electrons, $\Delta_{\mathrm{occ}} \sim 10$ eV.
A typical large basis set such as plane waves corresponds to $\Delta \sim 1 \, 000$ eV.
For the rational approximations in Eq. (\ref{minimax}), this corresponds to $y \sim 1 \, 000$.
For other rational approximations \cite{Lin_contour} on $[-y,y]$, this corresponds to $y \sim 100 \, 000$.
Before including other factors, this change in $y$ corresponds to a $\approx 50\%$ increase in the number of poles.
A less typical and more extreme electronic structure application is warm dense matter \cite{warm_dense_matter}
 with $\beta^{-1} \sim 0.1-100$ eV.
Eq. (\ref{empirical_error}) is invalid in the extreme case of $y \sim 1$,
 but $\epsilon_n(y)$ still decays exponentially in $n$ with a more complicated $y$-dependence.

\begin{figure}[t]
\includegraphics{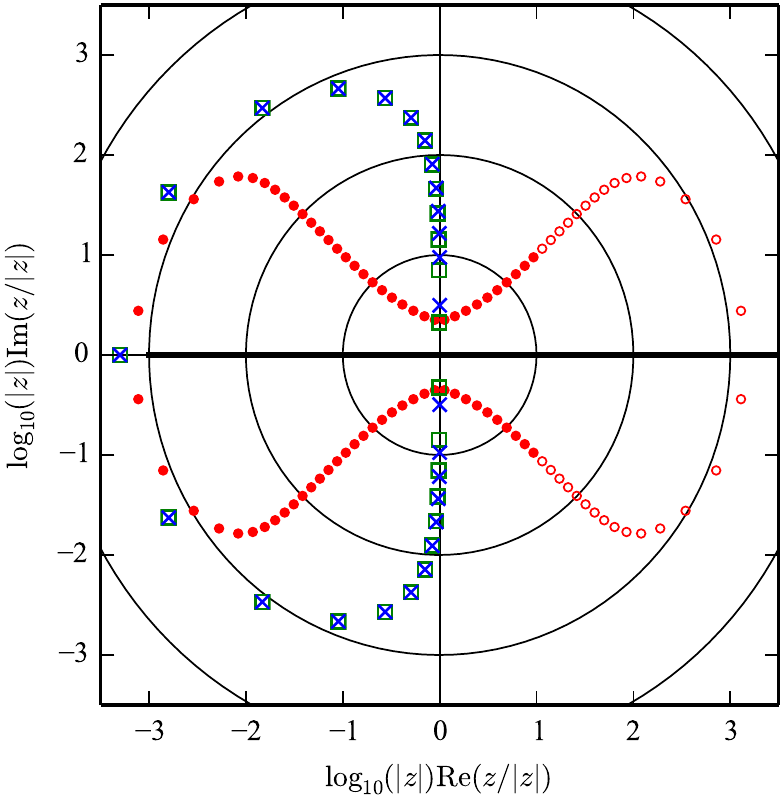}
\caption{\label{fig_poles} Complex pole structure of three rational approximations of the Fermi-Dirac distribution on $[-y,\infty)$ for $y =$ 1,000.
Previous work \cite{Lin_contour} using contour integral quadrature (circles) requires $n = 100$ poles to reduce the error to $\epsilon \approx 1.2 \times 10^{-7}$.
Of these poles, 30 have residues that are negligible in magnitude (open circles) and can be discarded.
The minimax approximation (crosses) requires only $n = 25$ poles to reduce the error to a similar value of $\epsilon \approx 4.2 \times 10^{-8}$.
The minimax approximation of $\theta(x)$ (squares) with the same value of $n$, $y$, and $\epsilon$ for $\delta = 7.52$ has nearly identical pole locations far from the origin.
The minimax and quadrature approximations are qualitatively different.
}
\end{figure}

The maximum pointwise error of an approximation to the Fermi-Dirac distribution
 can be related directly to the error in expectation values.
For two Hermitian matrices $H$ and $X$ that represent a mean-field electronic Hamiltonian and observable,
 the equilibrium expectation value of $X$ at temperature $\beta^{-1}$ is
\begin{equation} \label{expectation_value}
 \langle X \rangle_\beta = \mathrm{tr}[X (1 + \exp(\beta(H-E_F)))^{-1}] .
\end{equation}
We can approximate this expectation value using only shifted matrix inverses of $H$ and solutions of Eq. (\ref{minimax}) for $y = \beta \Delta_{\mathrm{occ}}$ as
\begin{equation} \label{expectation_approximation}
 \langle \widetilde{X} \rangle_\beta = \sum_{i=1}^n w_i \mathrm{tr}[X (\beta(H-E_F) - z_i)^{-1}] .
\end{equation}
The only dependence on $H$ is through its minimum eigenvalue $E_{\min}$ in $\Delta_{\mathrm{occ}} = E_F - E_{\min}$.
Using standard trace inequalities, we can bound the expectation value error in $\langle \widetilde{X} \rangle$ as
\begin{equation} \label{observable_error}
 \left| \langle X \rangle_\beta - \langle \widetilde{X} \rangle_\beta \right| \le \| X \|_1 \epsilon_n(\beta \Delta_{\mathrm{occ}}),
\end{equation}
 where $\| X \|_1$ is the sum of the singular values of $X$.
This bound does not depend directly on the matrix dimension of $H$ or any details of its energy spectrum other than $E_{\min}$.
These are the benefits of using the maximum pointwise error in Eq. (\ref{minimax}).
As with many error bounds in numerical linear algebra, Eq. (\ref{observable_error}) will be a very loose bound in typical calculations
 where there are many opportunities for cancellations of errors to occur.

\begin{figure*}[t]
\includegraphics{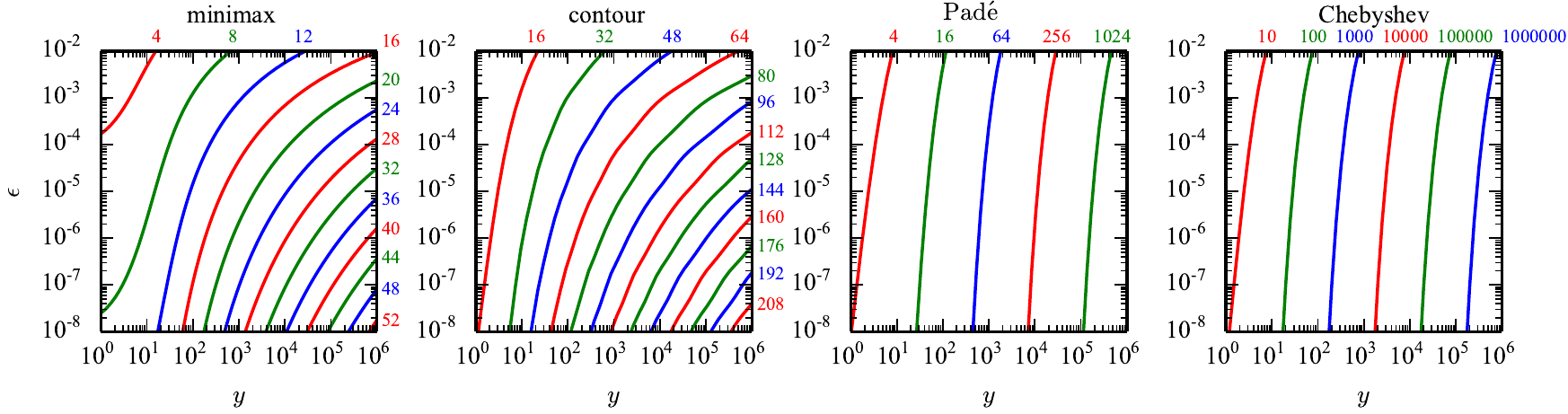}
\caption{\label{fig_compare} Maximum error $\epsilon$ of
 minimax rational approximation (far left) and contour integration quadrature \cite{Lin_contour} (left) on $[-y,\infty]$ and
 Pad\'{e} approximation \cite{pade_eigen} (right) and Chebyshev polynomial expansion (far right) on $[-y,y]$.
Labels denote number of poles or polynomial degree.
}
\end{figure*}

We compare the minimax approximation with the previous best approximation from contour integration quadrature \cite{Lin_contour} in Fig. \ref{fig_poles}.
In the previous work, the Fermi-Dirac distribution was split into $\tfrac{1}{2} - \tfrac{1}{2} \tanh(x)$, and $\tanh(x)$ was approximated with a rational function.
This splitting produces an approximation for $x \in [-y,y]$ that is not asymptotically exact for $x \gg 1$.
If the Fermi-Dirac distribution is instead approximated directly, then the approximation is valid for $x \in [-y,\infty)$.
Also, many of the residues on the positive real half-plane will be vanishingly small, decreasing the number of relevant poles by up to half.
The minimax approximation uses $\approx 75 \%$ fewer poles without removing poles with small residues as shown in Fig. \ref{fig_compare},
 and it uses $\ge 50 \%$ fewer poles even when poles are discarded.
Also, the quadrature construction restricts $n$ to a multiple of four.

The development of a new analytic rational approximation of the Fermi-Dirac distribution
 that is close in accuracy to the minimax approximation is an interesting open problem.
The existing quadrature result is an application of a more general conformal mapping technique \cite{contour_theory}.
This technique is able to reproduce the minimax rational approximation of $f(x) = \sqrt{x}$,
 which transforms into the minimax rational approximation of $\mathrm{sgn}(x)$ as $f(x^2)/x$.
These results rely on $f(x)$ having a simple branch cut along the negative real axis,
 and do not extend to $\sqrt{x} \tanh(\sqrt{x})$ and $\tanh(x)$ because they introduce poles along the branch cut.
Distinct contours are required for $\mathrm{sgn}(x)$ and $\tanh(\lambda x)$ even as they converge in the $\lambda \rightarrow \infty$ limit.

An alternative analytic approach to rational approximation of the Fermi-Dirac distribution
 is to coarsen poles far from the real axis.
The conventional Matsubara expansion is a simple truncation of distant poles,
 and it has been improved through the use of continued fractions \cite{continued_fraction}, partial fractions \cite{partial_fraction},
 and Pad\'{e} approximation \cite{pade_approx,pade_eigen}.
In Fig. \ref{fig_compare}, a representative Pad\'{e} approximation surpasses a Chebyshev polynomial expansion,
 but it is inferior to contour integration quadrature for large $y$.
Because these are fundamentally local approximations about the origin,
 the number of poles at fixed error scales as $O(\sqrt{y})$ instead of $O(\log y)$.
An $O(\log y)$ scaling can be recovered with coarsenings inspired by the fast multipole method \cite{Lin_multipole}.
These approximations typically restrict poles to the imaginary axis when approximating the symmetric $\tanh(x)$,
 but poles tend to be drawn into the negative real half-plane when capturing the asymptotic decay of the Fermi-Dirac distribution as $x \rightarrow \infty$.

With an optimal rational approximation of the Fermi-Dirac distribution,
 we lose some flexibility.
Approximations using a discretized contour integral around the spectrum of $H$ \cite{Lin_contour},
\begin{equation} \label{contour_fit}
 f(H) = \frac{1}{2 \pi i} \oint \frac{f(z) dz}{z - H} \approx \sum_{i=1}^n \frac{f(z_i) D_i}{z_i - H}
\end{equation}
 can be adapted to any function $f(x)$ that is analytic within the contour
 without recomputing expensive matrix inverses of $H$.
Without this structure, we can still reuse matrix inverses of $H$ by reoptimizing residues for each $f(x)$.
Alternatively, we can target multiple $f(x)$ with distinct error bounds as in Eq. (\ref{observable_error})
 and fit a common set of poles to minimize the maximum error in all bounds simultaneously.
This more complicated process can still utilize the methods established in this paper.

The most prevalent $f(H)$ in electronic structure applications is the single-electron density matrix
 in Eq. (\ref{expectation_value}) where $f(x)$ is the Fermi-Dirac distribution.
The electronic energy $\langle H - E_F \rangle_\beta$ contains a formally distinct $f(H)$,
 but we can calculate it using Eq. (\ref{expectation_approximation}) without extra matrix inversions.
The electronic free energy $\Omega_\beta(H - E_F)$ is a nontrivial distinct $f(H)$,
\begin{equation} \label{free_energy}
 \Omega_\beta(H - E_F) = - \beta^{-1} \mathrm{tr} [\ln( 1 + \exp( \beta (E_F - H)))] .
\end{equation}
This function has similar analytic structure to the Fermi-Dirac distribution,
 with branch points rather than simple poles along the imaginary axis.
We can relate expectation values to linear response properties of $\Omega_\beta(H - E_F)$,
\begin{equation}
 \langle X \rangle_\beta = \left( \frac{d}{d\lambda}\Omega_\beta(H - E_F + \lambda X) \right|_{\lambda = 0} .
\end{equation}
For $X = H - E_F$, this formula relates $\Omega_\beta(H - E_F)$ to an integral over Fermi-Dirac distributions with varying $\beta$,
\begin{equation}
 \Omega_\beta(H - E_F) = \lim_{\delta \rightarrow 0} \left( \Omega_\delta(H - E_F) + \int_\delta^\beta \langle H - E_F \rangle_{\beta'} \frac{d\beta'}{\beta'} \right) .
\end{equation}
Formally, we can fit a rational approximation of $\Omega_\beta(H - E_F)$ by summing over
 fits of Fermi-Dirac distributions for multiple $\beta$ values.
Practically, the $\delta \rightarrow 0$ limit has cancelling divergences between the two terms that cause numerical problems.

 Another source of $f(H)$ are fast algorithms for many-body perturbation theory.
These functions are produced by splitting energy denominators in multi-electron operators such as
\begin{equation}
 M = \frac{P_> \otimes P_<}{ H \otimes I - I \otimes H + \omega} ,
\end{equation}
 where $P_<$ is a projector onto the occupied subspace of $H$ with eigenvalues less than $E_F$
 and $P_>$ is a projector onto the virtual subspace of $H$ with eigenvalues greater than $E_F$.
The standard method in quantum chemistry is the Laplace transform \cite{Laplace_transform},
\begin{equation}
 M = \int_0^{\infty} e^{-\omega s} \left( e^{ (E_F-H) s} P_> \right) \otimes \left( e^{(H-E_F) s} P_< \right) ds ,
\end{equation}
but its projected exponentials are not analytic functions of $H$.
We can alternatively split $M$ using contour integration \cite{RPA_method},
\begin{align} \label{contour_split}
 M &= \frac{-1}{4 \pi^2} \oint_{<} \oint_{>} \frac{1}{z_> - z_< + \omega} \frac{dz_>}{z_>-H} \otimes \frac{dz_<}{z_<-H} ,
\end{align}
 to utilize analytic complex-shifted inverses of $H$ directly.

From numerical experiments, we find that poles optimized for the Fermi-Dirac distribution at a specific $\beta$ value
 degrade in accuracy when reused for other $\beta$ values or $\Omega_\beta(H - E_F)$ in Eq. (\ref{free_energy}).
This establishes the need for joint minimax rational approximation over multiple functions with a common set of poles.
For the relatively limited set of functions of $H$ that are needed in typical electronic structure applications such as in Eq. (\ref{contour_split}),
 we expect continued improvement over the use of Eq. (\ref{contour_fit}) with existing contour integration quadratures.

To demonstrate how minimax rational approximation of the Fermi-Dirac distribution can be useful for electronic structure applications,
 we apply it to a H\"{u}ckel model of polyacetylene.
In this canonical quantum chemistry problem, the $\pi$ electrons are described by a single atomic orbital per carbon atom with
 nearest-neighbor hopping described by a Hamiltonian matrix
\begin{equation}
 [H]_{i,j} = - t_i \delta_{i,j-1} - t_{j} \delta_{i-1,j}
\end{equation}
 with hopping energy $t_i$ between atoms $i$ and $i+1$.
The typical range of values is $2.6 \, \mathrm{eV} \le t_i \le 3.0 \, \mathrm{eV}$.
We calculate $\langle H \rangle_\beta$ for $\beta^{-1} = 0.03$ eV and $E_F = 0$.
Even for such a simple problem, we can highlight the advantages of rational approximation.

\begin{figure}[t]
\includegraphics{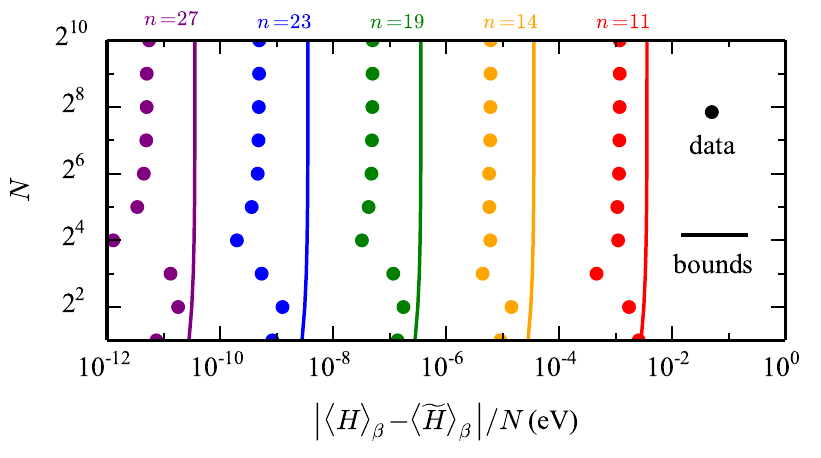}
\caption{\label{fig_error} Errors in $\langle H \rangle_\beta$ for the Huckel model of polyacetylene
 with uniform $t = 2.8$ eV compared to the error bound in Eq. (\ref{observable_error}).
}
\end{figure}

For a uniform polyacetylene chain with $t_i = t$, there are well known analytic results \cite{huckel_solution}.
We can calculate $\langle H \rangle_\beta$ directly by summing over the eigenvalues $E_i$ of $H$,
\begin{equation} \label{eigen_huckel}
 \langle H \rangle_\beta = \sum_{i=1}^{N} \frac{E_i}{1 + \exp(\beta E_i)}, \ 
 E_i = - 2 t \cos\left(\frac{i \pi}{N+1} \right) ,
\end{equation}
 for a chain containing $N$ carbon atoms.
We can also calculate the single-electron Green's functions \cite{tridiag_inverse},
\begin{equation} \label{inverse_huckel}
 \left[ \frac{1}{\beta H - z} \right]_{i,j} = \frac{g_{\min\{i,j\}} g_{N+1-\max\{i,j\}}}{g_{N+1}} , \  g_i = \frac{1}{\beta t} U_{i-1}\left(\frac{-z}{2\beta t}\right) ,
\end{equation}
 where $U_i(x)$ are Chebyshev polynomials of the second kind,
 and use Eq. (\ref{expectation_approximation}) to approximate $\langle H \rangle_\beta$.
We compare the errors of this approximation with bounds from Eq. (\ref{observable_error}) in Fig. \ref{fig_error}.

For a nonuniform polyacetylene chain, we begin to observe important differences between
 eigenvalue decomposition and matrix inversion.
Analytic results are no longer available with either method.
However, the inverse of a general tridiagonal Hermitian matrix retains the semiseparable form in Eq. (\ref{inverse_huckel})
 with a more complicated $g_i$ vector \cite{tridiag_inverse2}.
Its eigenvalues and eigenvectors have hierarchical structure that can be exploited by fast algorithms \cite{tridiag_eigen},
 but both the structure and algorithms are significantly more complicated than for matrix inversion.
This result mirrors the general case of sparse matrices that are relevant to electronic structure applications in 3 dimensions.
There is active development of structured matrix formats that are approximately closed under inversion
 and able to encode sparse matrices \cite{matrix_structure}.
These methods have not yet developed enough to be used in electronic structure applications,
 and the corresponding structured eigensolvers are just emerging \cite{eigen_structure}.
Electronic structure could benefit greatly from advances in the numerical linear algebra of sparse and structured matrices
 as it is often the bottleneck in electronic structure calculations.

In conclusion, we have calculated and tabulated minimax rational approximations of the Fermi-Dirac distribution
 that reduce cost prefactors in many sub-cubic-scaling algorithms for electronic structure \cite{quadratic_scaling,linear_scaling}.
Use of these algorithms over conventional cubic-scaling algorithms is limited by large cost prefactors,
 and every cost reduction is one step closer to their widespread adoption.
This result has benefitted from being a well-encapsulated function approximation problem
 with clear tradeoffs between cost and accuracy.
We should seek similar clarity in the relationships between cost and accuracy for the electron correlation models,
 basis sets, and numerical linear algebra that are relevant to electronic structure calculations
 in order to reduce their cost and improve their accuracy.

\bigskip
I thank Andrew Baczewski and Toby Jacobson for discussions and proofreading.
This work was supported by the NNSA Advanced Simulation and Computing -- 
 Physics and Engineering Models program at Sandia National Laboratories.
Sandia National Laboratories is a multi-program laboratory managed and 
operated by Sandia Corporation, a wholly owned subsidiary of Lockheed 
Martin Corporation, for the U.S. Department of Energy's National  
Nuclear Security Administration under contract DE-AC04-94AL85000.


\begin{thebibliography}{11}
\bibitem{quadratic_scaling} L. Lin, M. Chen, C. Yang, and L. He, \href{http://dx.doi.org/10.1088/0953-8984/25/29/295501}{J. Phys. Condens. Matter \textbf{25}, 295501 (2013)}.
\bibitem{linear_scaling} S. Goedecker, \href{http://dx.doi.org/10.1103/RevModPhys.71.1085}{Rev. Mod. Phys. \textbf{71}, 1085 (1999)}.
\bibitem{GW_method} C. Rostgaard, K. W. Jacobsen, and K. S. Thygesen, \href{http://dx.doi.org/10.1103/PhysRevB.81.085103}{Phys. Rev. B \textbf{81}, 085103 (2010)}.
\bibitem{RPA_method} J. E. Moussa, \href{http://dx.doi.org/10.1063/1.4855255}{J. Chem. Phys. \textbf{140}, 014107 (2014)}.
\bibitem{RPA_method2} M. Kaltak, J. Klime\v{s}, and G. Kresse, \href{http://dx.doi.org/10.1021/ct5001268}{J. Chem. Theory Comput. \textbf{10}, 2498 (2014)}.
\bibitem{Goedecker_contour} S. Goedecker, \href{http://dx.doi.org/10.1103/PhysRevB.48.17573}{Phys. Rev. B \textbf{48}, 17573 (1993)}.
\bibitem{Lin_contour} L. Lin, J. Lu, L. Ying, and E. Weinan, \href{http://dx.doi.org/10.1007/s11401-009-0201-7}{Chin. Ann. Math. Ser. B \textbf{30}, 729 (2009)}.
\bibitem{arXiv_supplemental_info} See ancillary files for source code and data tables.
\bibitem{Remez} W. Fraser and J. F. Hart, \href{http://dx.doi.org/10.1145/368273.368578}{Commun. ACM \textbf{5}, 401 (1962)}.
\bibitem{Zolotarev} T.-W. Chiu, T.-H. Hsieh, C.-H. Huang, and T.-R. Huang, \href{http://dx.doi.org/10.1103/PhysRevD.66.114502}{Phys. Rev. D \textbf{66}, 114502 (2002)}.
\bibitem{Cauchy_determinant} S. Schechter, \href{http://dx.doi.org/10.2307/2001955}{Math. Comp. \textbf{13}, 73 (1959)}.
\bibitem{warm_dense_matter} M. Koenig, A. Benuzzi-Mounaix, A. Ravasio, T. Vinci, N. Ozaki, S. Lepape, D. Batani, G. Huser, T. Hall, and D. Hicks, \href{http://dx.doi.org/10.1088/0741-3335/47/12B/S31}{Plasma Phys. Control. Fusion \textbf{47}, B441 (2005)}.
\bibitem{contour_theory} N. Hale, N. J. Higham, and L. N. Trefethen, \href{http://dx.doi.org/10.1137/070700607}{SIAM J. Numer. Anal. \textbf{46}, 2505 (2008)}.
\bibitem{continued_fraction} T. Ozaki, \href{http://dx.doi.org/10.1103/PhysRevB.75.035123}{Phys. Rev. B \textbf{75}, 035123 (2007)}.
\bibitem{partial_fraction} A. Croy and U. Saalmann, \href{http://dx.doi.org/10.1103/PhysRevB.80.073102}{Phys. Rev. B \textbf{80}, 073102 (2009)}.
\bibitem{pade_approx} J. Hu, R.-X. Xu, and Y. Yan, \href{http://dx.doi.org/10.1063/1.3484491}{J. Chem. Phys. \textbf{133}, 101106 (2010)}.
\bibitem{pade_eigen} C. Karrasch, V. Meden, and K. Sch\"{o}nhammer, \href{http://dx.doi.org/10.1103/PhysRevB.82.125114}{Phys. Rev. B \textbf{82}, 125114 (2010)}.
\bibitem{Lin_multipole} L. Lin, J. Lu, R. Car, and W. E, \href{http://dx.doi.org/10.1103/PhysRevB.79.115133}{Phys. Rev. B \textbf{79}, 115133 (2009)}.
\bibitem{Laplace_transform} J. Alml\"{o}f, \href{http://dx.doi.org/10.1016/0009-2614(91)80078-C}{Chem. Phys. Lett. \textbf{181}, 319 (1991)}.
\bibitem{huckel_solution} A. A. Frost and B. Musulin, \href{http://dx.doi.org/10.1063/1.1698970}{J. Chem. Phys. \textbf{21}, 572 (1953)}.
\bibitem{tridiag_inverse} D. Kershaw, \href{http://dx.doi.org/10.1090/S0025-5718-1969-0238478-X}{Math. Comp. \textbf{23}, 189 (1969)}.
\bibitem{tridiag_inverse2} P. Schlegel, \href{http://dx.doi.org/10.1090/S0025-5718-1970-0273798-2}{Math. Comp. \textbf{24}, 665 (1970)}.
\bibitem{tridiag_eigen} M. Gu and S. C. Eisenstat, \href{http://dx.doi.org/10.1137/S0895479892241287}{SIAM J. Matrix Anal. Appl. \textbf{16}, 172 (1995)}.
\bibitem{matrix_structure} S. B\"{o}rm, L. Grasedyck, and W. Hackbusch, \href{http://dx.doi.org/10.1016/S0955-7997(02)00152-2}{Eng. Anal. Bound. Elem. \textbf{27}, 405 (2003)}.
\bibitem{eigen_structure} J. Vogel, J. Xia, S. Cauley, and V. Balakrishnan, \href{http://dx.doi.org/10.1137/15M1018812}{SIAM J. Sci. Comput. \textbf{38}, A1358 (2016)}.
\end{thebibliography}
\end{document}